\documentclass[sigconf,final]{acmart}
\AtBeginDocument{%
  \providecommand\BibTeX{{%
    \normalfont B\kern-0.5em{\scshape i\kern-0.25em b}\kern-0.8em\TeX}}}

\usepackage{subfigure}
\usepackage{graphicx, caption}

\newcommand{\revise}[1]{{\color{black} #1}}

\begin{document}

%%
%% The "title" command has an optional parameter,
%% allowing the author to define a "short title" to be used in page headers.
\title{LPFS: Learnable Polarizing Feature Selection for Click-Through Rate Prediction}

%%
%% The "author" command and its associated commands are used to define
%% the authors and their affiliations.
%% Of note is the shared affiliation of the first two authors, and the
%% "authornote" and "authornotemark" commands
%% used to denote shared contribution to the research.
\author{Yi Guo}
\authornote{Both authors contributed equally to this work.}
\email{yguocuhk@gmail.com}
\affiliation{%
  \institution{KuaiShou Technology}
  \city{Beijing}
  \country{China}
}

\author{Zhaocheng Liu}
\authornotemark[1]
\email{lio.h.zen@gmail.com}
\affiliation{%
  \institution{KuaiShou Technology}
  \city{Beijing}
  \country{China}
}
\author{Jianchao Tan}
\email{jianchaotan@kuaishou.com}
\affiliation{%
  \institution{KuaiShou Technology}
  \city{Beijing}
  \country{China}
}

\author{Chao Liao}
\email{liaochao@kuaishou.com}
\affiliation{%
  \institution{KuaiShou Technology}
  \city{Beijing}
  \country{China}
}
% \author{Daqing Chang}
% \email{changdaqing@kuaishou.com}
% \affiliation{%
%   \institution{KuaiShou Technology}
%   \city{Beijing}
%   \country{China}
% }

% \author{Qiang Liu}
% \email{qiang.liu@nlpr.ia.ac.cn}
% \affiliation{%
%   \institution{Chinese Academy of Sciences}
%   \city{Beijing}
%   \country{China}
% }
\author{Sen Yang}
\email{senyang.nlpr@gmail.com}
\affiliation{%
  \institution{KuaiShou Technology}
  \city{Beijing}
  \country{China}
}

\author{Lei Yuan}
\email{lyuan0388@gmail.com}
\affiliation{%
  \institution{KuaiShou Technology}
  \city{Beijing}
  \country{China}
}

\author{Dongying Kong}
\email{kongdongying@kuaishou.com}
\affiliation{%
  \institution{KuaiShou Technology}
  \city{Beijing}
  \country{China}
}
\author{Zhi Chen}
\email{chenzhi07@kuaishou.com}
\affiliation{%
  \institution{KuaiShou Technology}
  \city{Beijing}
  \country{China}
}
% \author{Chengru Song}
% \email{songchengru@kuaishou.com}
% \affiliation{%
%   \institution{KuaiShou Technology}
%   \city{Beijing}
%   \country{China}
% }

\author{Ji Liu}
\email{ji.liu.uwisc@gmail.com}
\affiliation{%
  \institution{KuaiShou Technology}
  \city{Beijing}
  \country{China}
}

%%
%% By default, the full list of authors will be used in the page
%% headers. Often, this list is too long, and will overlap
%% other information printed in the page headers. This command allows
%% the author to define a more concise list
%% of authors' names for this purpose.
%\renewcommand{\shortauthors}{Trovato and Tobin, et al.}

%%
%% The abstract is a short summary of the work to be presented in the
%% article.
\begin{abstract}

In industry, feature selection is a standard but necessary step to search for an optimal set of informative feature fields for efficient and effective training of deep Click-Through Rate (CTR) models. Most previous works measure the importance of feature fields by using their corresponding continuous weights from the model, then remove the feature fields with small weight values. However, removing many features that correspond to small but not exact zero weights will inevitably hurt model performance and not be friendly to hot-start model training. There is also no theoretical guarantee that the magnitude of weights can represent the importance, thus possibly leading to sub-optimal results if using these methods.

To tackle this problem, we propose a novel Learnable Polarizing Feature Selection (LPFS) method using a smoothed-$\ell^0$ function in literature. Furthermore, we extend LPFS to LPFS++ by our newly designed smoothed-$\ell^0$-liked function to select a more informative subset of features. LPFS and LPFS++ can be used as gates inserted at the input of the deep network to control the active and inactive state of each feature. When training is finished, some gates are \textbf{exact zero}, while others are around one, which is particularly favored by the practical hot-start training in the industry, due to no damage to the model performance before and after removing the features corresponding to \textbf{exact-zero} gates. Experiments show that our methods outperform others by a clear margin, and have achieved great A/B test results in KuaiShou Technology.

\end{abstract}

%%
%% The code below is generated by the tool at http://dl.acm.org/ccs.cfm.
%% Please copy and paste the code instead of the example below.
%%
% \begin{CCSXML}
% <ccs2012>
%  <concept>
%   <concept_id>10010520.10010553.10010562</concept_id>
%   <concept_desc>Computer systems organization~Embedded systems</concept_desc>
%   <concept_significance>500</concept_significance>
%  </concept>
%  <concept>
%   <concept_id>10010520.10010575.10010755</concept_id>
%   <concept_desc>Computer systems organization~Redundancy</concept_desc>
%   <concept_significance>300</concept_significance>
%  </concept>
%  <concept>
%   <concept_id>10010520.10010553.10010554</concept_id>
%   <concept_desc>Computer systems organization~Robotics</concept_desc>
%   <concept_significance>100</concept_significance>
%  </concept>
%  <concept>
%   <concept_id>10003033.10003083.10003095</concept_id>
%   <concept_desc>Networks~Network reliability</concept_desc>
%   <concept_significance>100</concept_significance>
%  </concept>
% </ccs2012>
% \end{CCSXML}

% \ccsdesc[500]{Information systems~Online advertising}
% \ccsdesc[500]{Information systems~Data mining}

%%
%% Keywords. The author(s) should pick words that accurately describe
%% the work being presented. Separate the keywords with commas.
\keywords{Click-Through Rate Prediction, Feature Selection, Smoothed-L0}

%%
%% This command processes the author and affiliation and title
%% information and builds the first part of the formatted document.
\maketitle

\section{Introduction}
Click-Through Rate (CTR) prediction, which aims to estimate the probability of a user clicking on an item, has become a crucial task in industrial applications, such as personalized recommendations and online advertising~\cite{richardson2007predicting,zhou2018deep,zhou2019deep}.
In recent years, significant progress has been made due to the development of deep learning~\cite{guo2017deepfm,lian2018xdeepfm,yu2020deep,song2019autoint,fei2021gemnn},
however, these deep models still require an effective set of feature fields as input.
In industry, to accurately characterize user preferences, item characteristics, and contextual environments from different aspects, extensive feature engineering is generally essential for model training, which results in hundreds of feature fields in real-world datasets.

\begin{figure}
\centering
\subfigure[Distribution of weights norm for LASSO]{
\begin{minipage}[t]{0.23\textwidth}
\centering
\includegraphics[width = 1.0\textwidth]{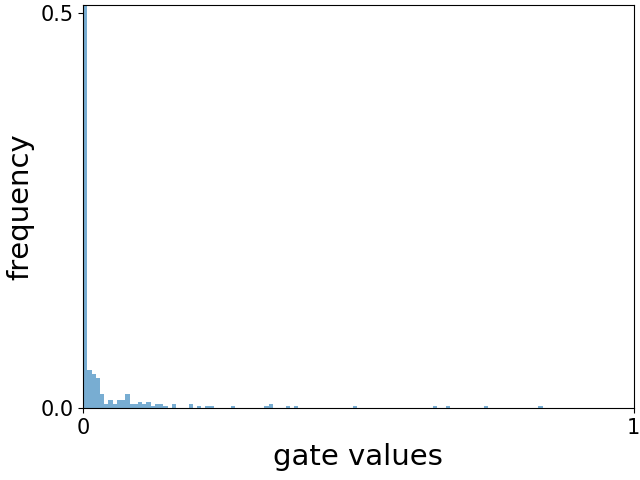}
\end{minipage}
}\subfigure[Distribution of gate values for LPFS]{
\begin{minipage}[t]{0.23\textwidth}
\centering
\includegraphics[width = 1.0\textwidth]{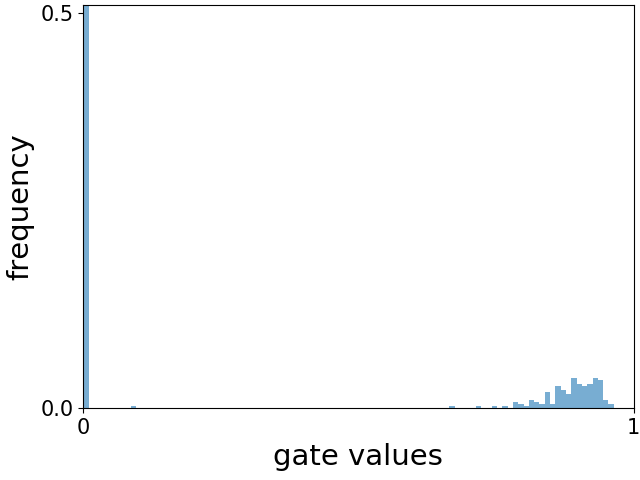}
\end{minipage}
}
\caption{The density distribution for group LASSO with proximal-SGD optimization (left) and for LPFS. The group LASSO method is implemented on the first layer of the network, and proximal SGD is applied. %We use proximal SGD here instead of the sub-gradient method to ensure that there is an interval that can be optimized to exactly zero. 
LASSO method outputs a continuous distribution and we need to choose a small threshold to determine whether to remove or keep the features. While our LPFS method outputs a polarized distribution: some gate values are exact zeros, the others are distributed around 1. We can remove the feature fields with exact zero gates, and absorb the non-zero-gates into the embedding or the weight of the first fully connected layer, so as to cause no damage to the model and be friendly to hot-start training. Although there are also exact-zero gates using LASSO, the value of many non-zero gates are very close to zero.}
\label{fig:distribution}
\end{figure}

However, we can only feed a subset of feature fields instead of all feature fields into models for effective and efficient training.
On the one hand, prior work \cite{shen2020stable,bahng2020learning} points out that the subtle dependencies among relevant features and irrelevant features may inflate the error of parameter estimation, resulting in instability of prediction results.
On the other hand, the online feature generation process consumes significant computing and storage resources.
Therefore how to search for an optimal set of effective feature fields from the real-world dataset is the core concern of both academia and industry, considering both effectiveness and efficiency.
Many feature selection methods have emerged in the past few decades.
Some methods \cite{meier2008group,friedman2010note,chapelle2014simple,2019AutoCross,liu2020dnn2lr,liu2021mining} are proposed to do feature selection for the Logistic Regression (LR) model which is the classical CTR prediction model.
For modern deep learning-based CTR prediction models, learning-based feature selection approaches have attracted much attention.
Specifically, COLD \cite{wang2020cold} applies the Squeeze-and-Excitation (SE) block~\cite{hu2018squeeze} to get the importance weights of features and select the most suitable ones. 
And some prior work \cite{li2016deep} proposes to select informative features via LASSO, while FSCD \cite{ma2021towards} via Gumbel-Softmax-liked sampling method.
More recently, UMEC \cite{shen2020umec} treats feature selection as a constrained optimization problem.
Besides, permutation-based feature importance \cite{molnar2020interpretable} measures the increase in the prediction error of the model after we permuted the feature’s values, which also has been widely applied in real-world feature selection. Despite the significant progress made with these methods, some challenges demanding further explorations:

(1) \textbf{Getting rid of ad-hoc thresholding}.
In general, previous methods output a \emph{continuous distribution} of importance weights for feature fields to represent the feature importance.
However, permutation-based, gumbel-softmax-based, or SE-based approaches cannot output \textbf{exact-zero} importance weights. They all need to prune from a skewed yet continuous histogram of importance weights, which makes the selection of the pruning threshold critical yet ad-hoc, taking Figure~\ref{fig:distribution} for example.

(2) \textbf{Hot-start-friendly}.
Empirically, feature selection usually needs a large range of data to train and generate confident feature selection results.
To save training cost, the \emph{hot-start} training is widely adopted in practice, which aims to inherit from the trained model and reduce the number of repetitions for training.
%For CTR prediction, the feature embeddings are the most important model parameters to \emph{hot start}.
The methods mentioned above remove many features with small but not exact zero importance weights. This step inevitably causes damage to the model's performance. Even for LASSO with the proximal optimization method, which can output exact zero weights, the remaining weights are very small and very close to zeros, which is unfriendly to hot-start. Last but not least, there is no theoretical guarantee that the magnitude of weights can represent the importance, leading to a suboptimal feature subset selected by these methods.

Inspired by the study of smoothed-$\ell^0$ \cite{mohimani2008fast,eftekhari2009robust, mohimani2010sparse, xiang2019new} in compressed sensing, we consider feature selection as an optimization problem under $\ell^0$-norm constraint, %and introduce the smoothed-$\ell^0$ formulation which satisfies the above conditions, to feature selection for deep learning-based CTR prediction models.
and propose a novel Learnable Polarizing Feature Selection (LPFS) method to effectively select highly informative features. We insert such differentiable function-based gates between the embedding and the input layer of the network to control the active and inactive state of these features. When training is finished, some gate values are \textbf{exact zero}, while others are distributed around one (see Figure~\ref{fig:distribution}). Then, we can remove feature fields with zero-gate, and absorb the non-zero gate to the embeddings or the weights of the first fully connected layer in the network. In this way, we can get rid of the threshold choosing, and there is totally no performance impact on the model performance before and after the physical removal, which is also very friendly to hot-start training.

Furthermore, each feature is unique, although maybe sometimes correlated, if a gate of a feature becomes zero accidentally, other features may not fully compensate for it. Then the model needs to have the ability to bring this feature to be active again. However, the derivative of all smoothed-$\ell^0$ functions proposed by previous works at $x=0$ is zero, which makes it impossible for gradient-based optimization methods to make a zero-gate become non-zero again. We are motivated to propose LPFS++ with a newly designed smoothed-$\ell^0$-liked function to alleviate this problem.
% and an extended version LPFS++.
% The proposed LPFS and LPFS++ get rid of ad-hoc thresholding, which makes them easy to use.
% To be noted, according to the smoothed L0 formulation of \cite{xiang2019new} which is used in LPFS, the gradient of a specified output weight is constantly zero when its value gets zero, which could be a problem when some outlier samples get the weight to zero accidentally. 
% The novel smoothed-L0-liked function in LPFS++ can alleviate such a problem, which is the main difference between LPFS and LPFS++.

%Specifically, in the early stage of training, if the weight of a specified feature field gets zero accidentally, this feature field will be discarded definitely.
%To alleviate such a problem, we propose LPFS++ which is based on a novel smoothed L0 formulation.

We conduct extensive experiments to verify the effectiveness of two proposed approaches on both public datasets and large-scale industrial datasets.
Experimental results demonstrate that LPFS++ outperforms LPFS, and both of them outperform previous approaches by a significant margin.
Our approaches have also been deployed on the offline Kuaishou distributed training platform and have been exploited in different scenarios in Kuaishou to do feature selection for their CTR prediction models and have achieved significant A/B testing results.

We summarize the contributions of the proposed methods below:
\begin{itemize}
\item We are pioneers to apply the idea of smoothed-$\ell^0$ gate to feature selection for CTR prediction and we propose a novel feature selection method by utilizing the closed-form proximal SGD updating for gate parameters, which outputs a polarized distribution of feature importance scores and is naturally friendly to hot-start training.
\item We extend the idea of smoothed-$\ell^0$ and propose LPFS++ which is based on a novel smoothed-$\ell^0$-liked function and can select features more robustly.
\item Experiments show that the both proposed methods outperform other feature selection methods by a clear margin, and they have achieved great benefits in KuaiShou Technology.
\end{itemize}

\section{RELATED WORKS}
\subsection{Smoothed-$\ell^0$ Optimization}
The idea of smoothed-$\ell^0$ optimization was first proposed in~\cite{mohimani2007fast,mohimani2008fast} to obtain sparse solutions for under-determined systems of linear equations. The original optimization objective is $\min_{x}\|x\|_0$ s.t. $Ax=y$, where $A\in \mathbb{R}^{m\times n}$, $x\in \mathbb{R}^n$, $y\in \mathbb{R}^m$ and $m < n$. This is intractable, and they turn around to optimize  $\min_{x}{ f_{\epsilon}(x)}=1-\exp{\left(\frac{-x^2}{2\epsilon^2}\right)}$ s.t. $Ax=y$.
%on the feasible set $x\in \{x|Ax=y\}$, where $A$ is a known matrix and $y$ is a known vector. 
When $\epsilon$ is large, $f_{\epsilon}(x)$ is smooth, while when $\epsilon$ is small enough, the function can approximate $\ell^0$-norm indefinitely. So, they solve it iteratively, gradually decaying $\epsilon$. And their experiments show that the method is much faster than $\ell^1$-based methods, while achieving the same or better accuracy. Following work~\cite{mohimani2010sparse} studied the convergence properties of the above smoothed-$\ell^0$ and find that under some mild constraints, the convergence is guaranteed. Afterwards, various smoothed-$\ell^0$ functions had been proposed and studied for compressed sensing, such as $f_{\epsilon}(x)=\sin\left(\arctan\left(\frac{|x|}{\epsilon} \right) \right)$\cite{wang2019re}, $f_{\epsilon}(x)=\tanh\left(\frac{x^2}{2\epsilon^2} \right)$~\cite{zhao2012reconstruction} and $f_{\epsilon}(x)=\frac{x^2}{x^2+\epsilon^2}$~\cite{xiang2019new}. All these functions have following important property:
\begin{equation}
    \lim_{\epsilon \rightarrow 0}{g_{\epsilon}(x)}=
    \begin{cases}
    0,\quad x=0\\
    1, \quad x \ne 0
    \end{cases}
    \label{related works}
\end{equation}
and can approximate $\ell^0$-norm well. In this paper, in addition to directly applying the existing smoothed-$\ell^0$ function as gate to control feature selection, we also propose a new designed smoothed-$\ell^0$-liked function for feature selection. We use the term "liked" here because what we proposed is an odd function, not an even function like all the above functions.
%\subsection{Click-Through Rate Prediction}
\subsection{Feature Selection}
Feature selection is a key component for CTR Prediction and various methods have been proposed in this area.
\cite{meier2008group,friedman2010note,chapelle2014simple,2019AutoCross,liu2020dnn2lr,liu2021mining} are proposed to do feature selection for the Logistic Regression (LR) model which is the classical CTR prediction model. In the area of deep learning, COLD~\cite{wang2020cold} use Squeeze-and-Excitation(SE) block~\cite{hu2018squeeze} to measure the importance of features, while FSCD~\cite{ma2021towards} use gumbel-softmax/sigmoid~\cite{jang2016categorical}. 
These two works both output a continuous importance distribution via softmax or sigmoid function, and the importance scores can not become exact zero.
LASSO is also a common feature selection method in the industry~\cite{li2016deep}. With the proximal SGD~\cite{nitanda2014stochastic} algorithm, the LASSO method can push a portion of parameters to be exact zeros, but their distribution is  also continuous. All these methods require choosing a threshold to truncate the distribution. Recently, ~\cite{shen2020umec} use the alternating direction method of multipliers(ADMM) optimization method to select features, but we find it a little hard to compress a very large feature set into very small ones. In addition to this learning-based approach, feature permutation~\cite{molnar2020interpretable} is a common method in the industry. It first trains a model with all the features, and then keeps other features unchanged, permutes the input data along each feature axis randomly one by one, and uses the magnitude of the drop in performance as the metric for the importance of the features. This greedy method is easy to implement and does not require parameter tuning, but it ignores the possible correlation between features, and some features may compensate for each other.

Besides the direct feature selection, there is also some works on the interaction between features, such as AutoCross~\cite{2019AutoCross}, Autofis~\cite{liu2020autofis}, Autoint~\cite{song2019autoint}. In essence, we can also regard the cross feature as a common feature, add it to the feature superset, and perform a general feature selection to discover which features should be interacted with. We will study it through an experiment in Section~\ref{Terabyte With Cross Features}.

\section{METHODS}
\subsection{Problem Formulation}
For most deep-learning-based CTR prediction models, the input data is unusually collected in continuous and categorical forms. The continuous part can be fed into the network directly, while the categorical part is often pre-processed by mapping each category into a dense representation space~\cite{covington2016deep}, also known as embeddings. The embeddings and continuous features are then fed into the deep network. For description simplicity, we only describe the feature selection of categorical features in this paper. Suppose there are $N$ feature fields concatenated as input $\boldsymbol{e}=[\boldsymbol{e}_1, \boldsymbol{e}_2, ..., \boldsymbol{e}_N]$, each $\boldsymbol{e}_i, i=1,2,...,N$ is an embedding vector for the $i$-th field, and we want to select a most informative subset from them. The network can be formulated as a function:
\begin{equation}
    y=f(w; \boldsymbol{e})
\end{equation}
where $w$ is all the network parameters, $y$ the output of the network.
To select features, we can insert a vector-valued function $\boldsymbol{g}(x)=[g(x_1), g(x_2), ..., g(x_N)]$ as gate before the input of the network, where each $x_i, i=1,2,...,N$ can be either a new introduced learnable scalar parameter, or the weights norm from first fully connected layer, or even embedding information in some works~\cite{li2016deep}. Then the input can be viewed as $\boldsymbol{\tilde{e}}=[g(x_1)\boldsymbol{e}_1, g(x_2)\boldsymbol{e}_2,..., g(x_N)\boldsymbol{e}_N]$. In this way, the network becomes:
\begin{equation}
    y=f(w; \boldsymbol{g}(x)\boldsymbol{e})
    \label{network}
\end{equation}
\begin{figure}
\centering
\begin{minipage}[t]{0.4\textwidth}
\centering
\includegraphics[width = 1.0\textwidth]{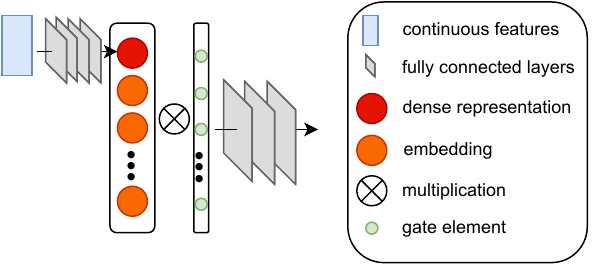}
\end{minipage}
\caption{This figure takes DLRM~\cite{naumov2019deep} on the Terabyte dataset as an example to show where we insert the gate vector. In this experiment, the 13 continuous features are transformed by a \textsl{dense}-MLP to a dense representation, we treat the dense representation as a feature for the subsequent feature selection. Then the dense representation and the 26 categorical embeddings are concatenated as 27 features. We multiply the 27 features by a gate vector with 27 elements, where each element control the liveness of each feature. The multiplication results are the new input to the \textsl{top}-MLP in DLRM.}
\label{dlrm}
\end{figure}
As shown in Figure~\ref{dlrm}, during training, the gates, embeddings, and all other network parameters are trained together, and we also impose some penalties on the gates parameters to implement feature selection. When training terminates, the features with zero-gates can be viewed as irrelevant features and can be removed, while others will be kept. 
Under this formulation, the key to feature selection becomes how to choose a good gate function. Previous works~\cite{li2016deep, ma2021towards} use lasso or gumble-softmax-trick~\cite{jang2016categorical} as the gate function, but these kinds of methods often output a continuous distribution of gates, then remove many feature fields with small but not exact zero gates. We don't think this is a good enough way and is unfriendly to hot-start training.

\subsection{LPFS}
\begin{figure}
\centering
\subfigure[$\ell^0$, $\ell^1$ and smoothed-$\ell^0$ $g_{\epsilon}(x)$ ]{
\begin{minipage}[t]{0.24\textwidth}
\centering
\includegraphics[width = 1.0\textwidth]{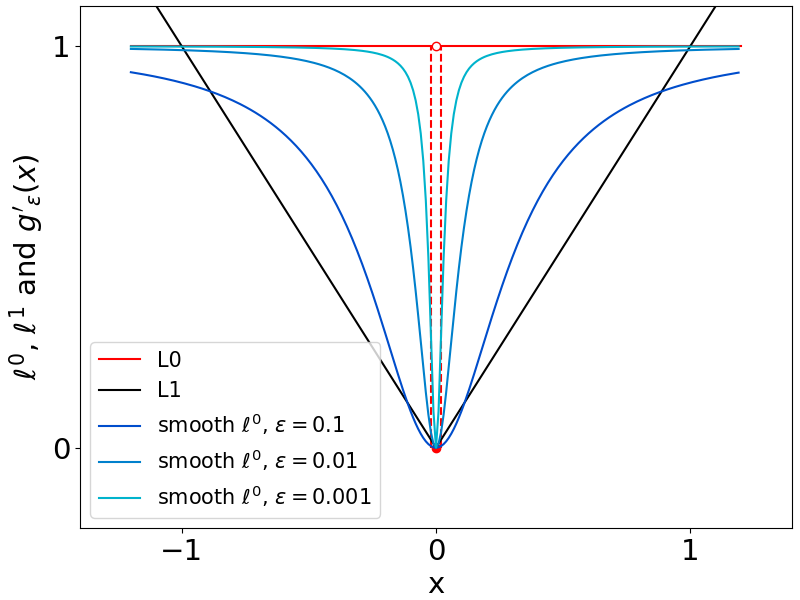}
\end{minipage}
}\subfigure[smoothed-$\ell^0$ $g'_{\epsilon}$ with different $\epsilon$]{
\begin{minipage}[t]{0.24\textwidth}
\centering
\includegraphics[width = 1.0\textwidth]{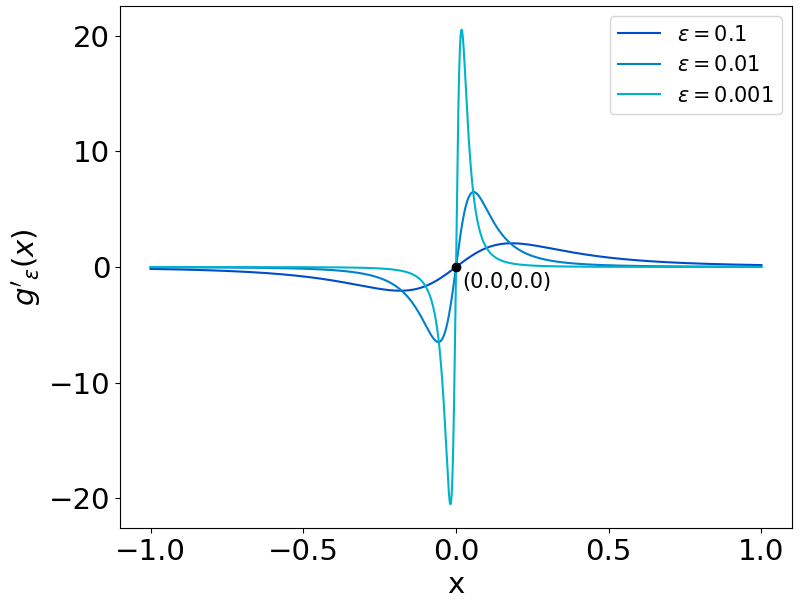}
\end{minipage}
}
\caption{The graph of $\ell^0$, $\ell^1$ and smoothed-$\ell^0$ $g_{\epsilon}(x)$ with different $\epsilon$ (left) and $g'_{\epsilon}$ (right). $g_{\epsilon}(x)$ a quasi-convex function between $\ell^0$ and $\ell^1$ function. When $\epsilon$ is large, $g_{\epsilon}(x)$ is smooth and continuously differentiable; while when $\epsilon$ is small enough, $g_{\epsilon}(x)$ can be infinitely close to $\ell^0$. As the $\epsilon$ approaching to zero, $g'_{\epsilon}(x)$ is zero only at the points where $g_{\epsilon}(x)$ get exact zeros or near ones, otherwise infinity. Note that the $g'_{\epsilon}(x)|_{x=0.0}$ is exactly zero whatever $\epsilon$ is.}
\label{fig: g and gd}
\end{figure}

From the perspective of optimization, feature selection can be essentially formulated as an optimization problem under $\ell^0$-norm constraint. Since it's an NP-hard to solve directly, some previous works turn around it by relaxing $\ell^0$ to $\ell^1$, which is a convex function closest to $\ell^0$. Inspired by the research field of smoothed-$\ell^0$ optimization, we use the following smoothed-$\ell^0$ function~\cite{xiang2019new} as our gate for LPFS:
\begin{equation}
    g_{\epsilon}(x) = \frac{x^2}{x^2+\epsilon}
    \label{smoothL0}
\end{equation}
This function has the following roperty:
\begin{equation}
    g_{\epsilon}(x)
    \begin{cases}
    =0,\quad x=0\\
    \approx 1, \quad x \ne 0
    \end{cases}
    \label{eq:property}
\end{equation}
The derivative of $g_{\epsilon}(x)$ w.r.t $x$ is:
\begin{equation}
    g'_{\epsilon}(x) = \frac{2x\epsilon}{(x^2+\epsilon)^2}
    \label{daoshu}
\end{equation}
where we choose $x$ here as newly introduced learnable parameters, $\epsilon$ is a small positive number. This is a quasi-convex function between $\ell^0$ and $\ell^1$ function, as shown in Figure \ref{fig: g and gd}. From the figure, we can see that when $\epsilon$ is large, $g_{\epsilon}(x)$ smooth and continuously differentiable and that when $\epsilon$ decays small enough, $g_{\epsilon}(x)$ can be infinitely close to $\ell^0$. Different from Gumble-softmax/sigmoid with decreasing temperature, function~\eqref{smoothL0} can be exactly zero even when $\epsilon$ is not small enough. And different from LASSO with proximal algorithms, function~\eqref{smoothL0} outputs a polarized distribution with some gate values being exactly zeros, while others have large margins from zeros. These are two very nice properties for a hot start, as when we remove features with exactly zero gates, the performance of the model remains unchanged.

\subsection{LPFS++}
Although LPFS has achieved great performance both in open benchmarks and KuaiShou's online A/B tests (as shown in the experiment section), there is big room for improvement in the task of feature selection. GDP~\cite{guo2021gdp} applies the idea of smoothed-$\ell^0$ in the field of channel pruning in Computer Vision. But feature selection is much different from channel pruning in the vision in two ways:

1. Channel pruning is essentially searching for the combination of the number of neurons for each layer. It only cares about the number of neurons in a certain layer, rather than which neurons. However, feature selection is different. We need to obtain a subset of features.  In addition to caring about how many features this subset contains, it is more important to care about which features. 

2. In the industry, user behavior is changing slowly. While mainstream datasets, such as ImageNet~\cite{deng2009imagenet}, are static, user data flows dynamically. This phenomenon requires the model to be more robust to feature selection than to channel pruning.
%Just because a feature isn't important now doesn't mean it won't be important later. 

\begin{figure}
\centering
\subfigure[$\ell^0$ and $g_{\epsilon++}(x)$ ]{
\begin{minipage}[t]{0.24\textwidth}
\centering
\includegraphics[width = 1.0\textwidth]{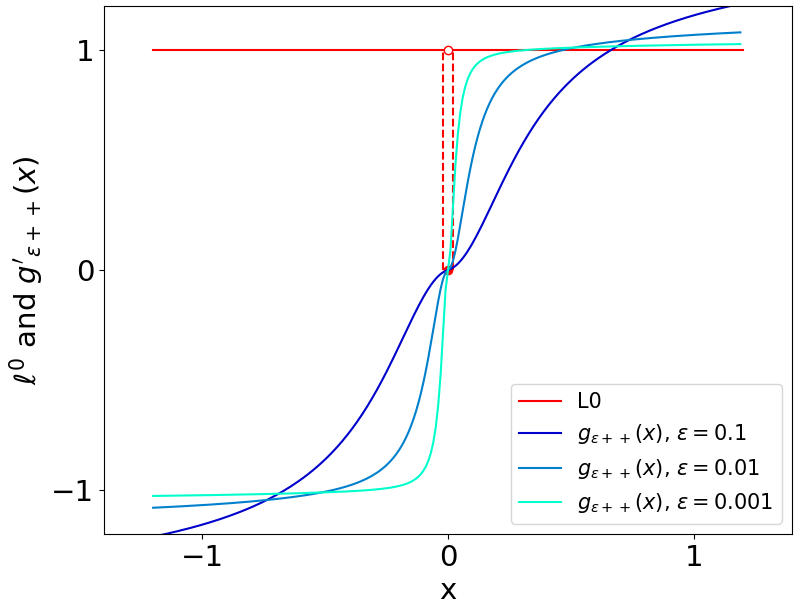}
\end{minipage}
}\subfigure[$g'_{\epsilon++}$ with different $\epsilon$]{
\begin{minipage}[t]{0.24\textwidth}
\centering
\includegraphics[width = 1.0\textwidth]{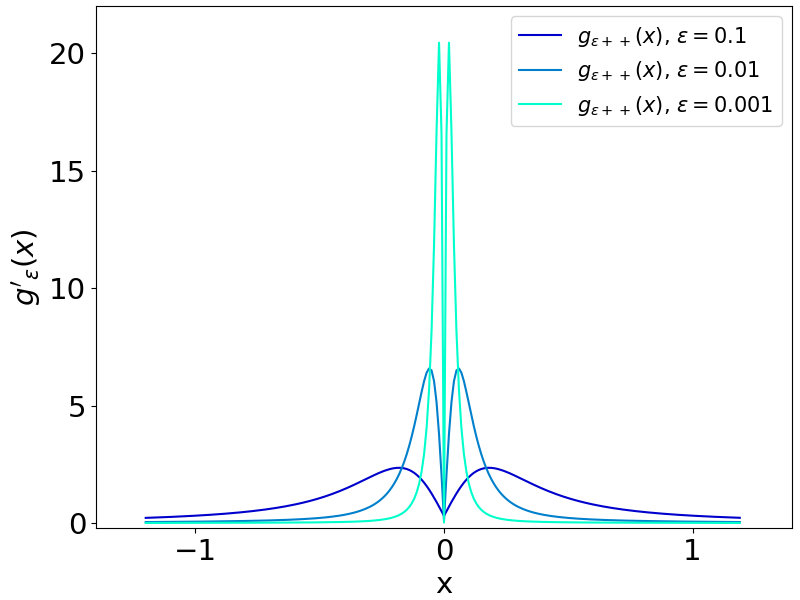}
\end{minipage}
}
\caption{The graph of $\ell^0$ and$g_{\epsilon++}(x)$ with different $\epsilon$ (left) and $g'_{\epsilon++}$ (right). Now $g_{\epsilon++}(x)$ is odd, not even like the smoothed $\ell^0$ function. Note that the gradient at $x=0$ is no longer zero, but a small number decays with $\epsilon$. }
\label{fig: lpp}
\end{figure}

Mathematically, what the function ~\eqref{smoothL0} needs to be improved for feature selection is that its derivative at $x=0$ is $0$, as easy to see in Eq.~\eqref{daoshu}. In this situation, the derivative of the output $y$ of network w.r.t $x$, from Eq.~\eqref{network}, is
\begin{equation}
    \frac{\partial y}{\partial x}\Big |_{x=0}=f'_{2}(w; \boldsymbol{g}_{\epsilon}(x)\boldsymbol{e})\boldsymbol{e}g'_{\epsilon}(x)|_{x=0}
    \label{equal6}
\end{equation}
where $f'_{2}(a;b)=\frac{\partial f}{\partial b}$, the subscript $2$ mean partial derivative w.r.t the second term.
What the problem Eq.~\eqref{equal6} will lead to is that whether some outlier samples or change in user behavior causes a gate value to go to zero accidentally, the corresponding feature will never be resurrected based on gradient-based optimization methods, and it could not be compensated by other features. One solution is to add some fading random noise, but we want to solve this from the gate function itself. We need to construct a gate function that satisfies properties similar to the smoothed-$\ell^0$ function, and whose derivative is not zero at $x=0$. One heuristic optional solution is:

\begin{equation}
    g_{\epsilon++}(x) = 
    \begin{cases}
    \frac{x^2}{x^2+\epsilon} + \alpha {\epsilon}^{1/\tau} \arctan(x), \quad  x \ge 0\\
    - \frac{x^2}{x^2+\epsilon} + \alpha {\epsilon}^{1/\tau} \arctan(x), \quad  x < 0
    \end{cases}
    \label{eq:SL0pp}
\end{equation}
Although it is a piece-wise function, it is also smooth and continuously differentiable. The derivative of $g_{\epsilon}(x)$ w.r.t $x$ is
\begin{equation}
    g'_{\epsilon++}(x) = \frac{2|x|\epsilon}{(x^2+\epsilon)^2} + \frac{\alpha {\epsilon}^{1/\tau}}{x^2+1}
    \label{SL0ppgredient}
\end{equation}
Where $x$ and $\epsilon$ have the same meaning as Eq.~\eqref{smoothL0}, $\alpha$ is a constant hyper-parameter balancing the two terms. \revise{The exponential factor $1/\tau$ is to control the decay rate of $g'_{\epsilon++}(x=0)$ with respect to $\epsilon$. This variable is pretty robust, thus we have taken $\tau=2$ in all of previous private experiments in KuaiShou Technology, which worked very well. Also, the arctangent function is not essential, because what we need is just an odd function whose derivative at $x=0$ is not zero, and whose value tends to be constant for large x. There are many other functions that can satisfy this properties, and might work better, which could be the future work.}
%The sign of square root $\sqrt{\cdot}$ here is not essential, what we want is a term that decays with $\epsilon$, and the $\sqrt{\cdot}$ here just slows down the decay rate. We can even take the third or fourth root or remove the square root, and that would probably work better. We use square root here, just because we found that we have used it in all previous private experiments, and it works.
The graph is shown as Figure~\ref{fig: lpp}. Note that, unlike all smoothed-$\ell^0$ functions, Function ~\eqref{eq:SL0pp} is an odd function rather than an even function. This is a necessary consequence because the derivative of a continuously differentiable even function at $x=0$ must be 0. It is fine to use an odd function as the gate, because we can absorb the negative signs of gates values into the corresponding embeddings or the first fully connected layer of the network. That is, in Eq.~\eqref{network} the equality holds: $\boldsymbol{g}(x)\boldsymbol{e} = abs(\boldsymbol{g}(x)) sign(\boldsymbol{g}(x))\boldsymbol{e} = abs(\boldsymbol{g}(x)) \boldsymbol{e'}$, where $\boldsymbol{e'}=sign(\boldsymbol{g}(x))\boldsymbol{e}$ is the final embeddings for downstream tasks.
Now we have $g'_{\epsilon++}(x=0)=\alpha {\epsilon}^{1/\tau}$, which is small number decays with $\epsilon$. Note that $g'_{\epsilon++}(x=0)\ne 0$ does not mean $\frac{\partial y}{\partial x}\Big |_{x=0}\ne 0$ because of the existence of $f'_{2}(w; \boldsymbol{g}_{\epsilon}(x)\boldsymbol{e})$ in Eq.~\eqref{equal6}. During the offline feature selection, $\epsilon$ initializes a large value so that the $g'_{\epsilon++}(x=0)$ is robust enough to outlier samples and the slow change of user behavior. As training goes on, $g'_{\epsilon++}(x=0)$ also decays as $\epsilon$ decays, so that the feature superset can be stably divided into \revise{non-in}formative and informative subset when $\epsilon$ decays to be small enough.
%F%or online feature selection, we just need to make $\epsilon$ decay not too small, which we find works well, and will be our future work. In this paper, we focus on offline feature selection.
\subsection{Optimization}

For both LPFS and LPFS++, we optimize the following objective function:
\begin{equation}
    \min_{w,x}\mathcal{F}(w;x) = \mathcal{L}(\hat{y};f(w; \boldsymbol{g}(x)\boldsymbol{e}))+\lambda \|x\|_1
    \label{eq:objective}
\end{equation}
Where $f(w; \boldsymbol{g}(x)\boldsymbol{e}))$ represents the network, same to Eq.~\eqref{network}; $\hat{y}$ is the ground true(i.e. users click or not click the item); $\mathcal{L}(\cdot;\cdot)$ is the loss function between the ground truth and model prediction; $\|\cdot \|_1$ is $\ell^1$-norm; $\lambda$ is the balance factor.  $w$ and $\boldsymbol{e}$ in $\mathcal{L}(\hat{y};f(w; \boldsymbol{g}(x)\boldsymbol{e}))$ are updated by Adam~\cite{kingma2014adam} or Adagrad~\cite{duchi2011adaptive} optimizer, same to baseline. And $x$ is updated by proximal-SGD. \revise{It is equivalent to solve $\min_x\{\frac{1}{2\eta}||x-\tilde{x}_t||^2+\lambda\|x\|_1\}$, whose closed-form solution is:}
\begin{equation}
x^{t+1} =
    \begin{cases}
    \tilde{x}_t-\lambda\eta, \quad \tilde{x}_t \ge \lambda\eta \\
    0, \quad  \quad -\lambda\eta < \tilde{x}_t < \lambda\eta, \\ 
    \tilde{x}_t+\lambda\eta, \quad \tilde{x}_t \leq -\lambda\eta \\
    \end{cases}
\label{eq: closed-form prox-l1}
\end{equation}
Where $\eta$ is learning rate for $x$, $\tilde{x}_t$ is the value of $x$ after one step of updating $\mathcal{L}(\hat{y};f(w; \boldsymbol{g}(x)\boldsymbol{e}))$ with the Momentum optimizer. The core code for updating $x$ is shown in supplementary material.
\\
\revise{It is worth pointing out that we use $\ell^1$-norm instead of $\ell^0$-norm to regularize $x$ in Eq.\eqref{eq:objective}. 
In fact, the polarization effect ($\ell^0$ property) that we're looking for is derived from smoothed-$\ell^0$ gate in Eq.~\eqref{smoothL0}, not from $\ell^1$ regularization on $x$, the $\ell^1$ regularization is only to penalize $x$ to let smoothed-$\ell^0$ gate become polarized like $\ell^0$.

Moreover, for $\ell^0$ regularization, the problem $min_x\{\frac{1}{2\eta}||x-\tilde{x}_t||^2+\lambda\|x\|_0\}$ indeed has closed-form solutions:
\begin{equation}
x^{t+1} =
    \begin{cases}
    0,&|\tilde{x}_t| < \sqrt{2\lambda\eta} \\
    \tilde{x}_t,&|\tilde{x}_t| > \sqrt{2\lambda\eta} \\ 
    0 \text{ or } \tilde{x}_t,&|\tilde{x}_t| = \sqrt{2\lambda\eta} \\
    \end{cases}
\label{eq: l0 close form solution}
\end{equation}
but we find it is very sensitive to the initial value of $x$ in Eq.~\eqref{eq:objective} in our experiments. If we initialize $x$ to be greater than $\sqrt{2\lambda\eta}$ accidentally, it can hardly be penalized during the whole training process under the second case in Eq.~\eqref{eq: l0 close form solution}; otherwise it becomes zero very quickly under the first case in Eq.~\eqref{eq: l0 close form solution}. Additionally, in each SGD iteration, $\tilde{x}_t$ is not necessarily optimal so it does not need to be directly updated by Eq.~\eqref{eq: l0 close form solution} to be zero. While in Eq.~\eqref{eq: closed-form prox-l1}, $x$ will be penalized by a small step $\lambda\eta$ whatever its initial value is.}

\section{EXPERIMENTS}
In this section, we will demonstrate the superiority of our method through three experiments. We mainly describe the core part of the experiments here, and the training details such as hyper-parameter setting are left at the end of this paper.
\subsection{Datasets}
To show that our method can filter out the highly informative features, we conducted experiments on a large-scale dataset Criteo AI Labs Ad Terabyte dataset~\footnote{\url{https://labs.criteo.com/2013/12/download-terabyte-click-logs/}} and an industrial dataset in KuaiShou. For Terabyte dataset, it contains 26 categorical features and 13 continuous features. It contains about 4.4 billion click log samples over 24 days. Similar to DLRM~\cite{naumov2019deep}, for negative samples, we randomly select 12.5\% on each day. To be fair, for all the public methods and our method, We pre-train the model by "day $0\sim17$", and select features by "day $18\sim22$", in which process, all the model parameters, including newly introduced parameters for FSCD~\cite{ma2021towards} and our methods, are trained together for learnable methods. When the feature subsets are selected by these methods, the models are then trained from scratch by "day $0\sim22$" using this subset, then evaluated by "day 23" and the best AUC calculated in official DLRM~\cite{naumov2019deep} code~\footnote{\url{https://github.com/facebookresearch/dlrm}} is reported.

The industrial dataset is collected by Mobile Kuaishou App, we select 9 days for offline features selections, and take 10\% of the negative samples at random. 
All the positive and negative click log samples add up to nearly 1.2 billion over the 9 days. This dataset contains 250 user categorical feature fields, 46 item categorical feature fields, 96 combine categorical feature fields, and 25 continuous features. 
Similar to the configuration of Terabyte, we pre-train the model by "day $1\sim6$", and select features by "day $7\sim8$", in which process all the parameters are trained together. When obtaining feature subsets, the models are trained by cold start by "day $1\sim8$" using these subsets, then are evaluated by "day 9". 
\subsection{Network Settings}
For the public Terabyte dataset, we made some minor modifications to the DLRM~\cite{naumov2019deep} to serve as our baseline network. The 13 continuous features in DLRM are transformed by an MLP to a dense representation with the same length as embeddings. Then we treat this dense representation as an embedding, with the same status as the other 26 embedding features. In this way, we have 27 features in the feature superset. Since the 27 features are still very small and the subset of features selected by various methods is similar, we also conducted another experiment with all the crossed features in addition to this direct selection, which will be explained in the following subsections. Same to ~\cite{shen2020umec}, we set the hidden dimensions of the three-layer MLP prediction model as 256 and 128 for both crossed and non-crossed versions.

For the industrial dataset, every categorical feature is 16 dimension, while the different dense feature has different dimensions, ranging from 1 to 128.
 Every user feature field and every item feature field is crossed by element-wise multiplication.
 Then the crossed features, combine features, and continuous features are concatenated as the input for the network.
 The network contains a share bottom layer, then some auxiliary branches for different auxiliary tasks.
 We focus on one main branch. 

\subsection{Terabyte Without Cross Features}
In this experiment, we treat the 27 features(including a dense representation transformed by 13 dense features, as described in the last subsection) as a feature superset for feature selection. Each feature is 16 dimensions. These 27 features are multiplied by the corresponding gates, then concatenated as a vector, and fed into the three-layer MLP prediction branch (or called the \textsl{top}-MLP in the original paper), as shown in Figure~\ref{dlrm}. We compared our method with FSCD~\cite{ma2021towards}, UMEC~\cite{shen2020umec}, feature permutation, and group LASSO. For UMEC, all the hyper-parameters are set as the paper reported; for FSCD, we set all the regularization weights to the same values, for group LASSO, we regularize the weights of the first fully connected layer in the \textsl{top}-MLP, and train them by proximal-SGD; for feature permutation, we randomly shuffle the feature to be evaluated in a mini-batch. 

\begin{figure}
\centering
\begin{minipage}[t]{0.45\textwidth}
\centering
\includegraphics[width = 0.85\textwidth]{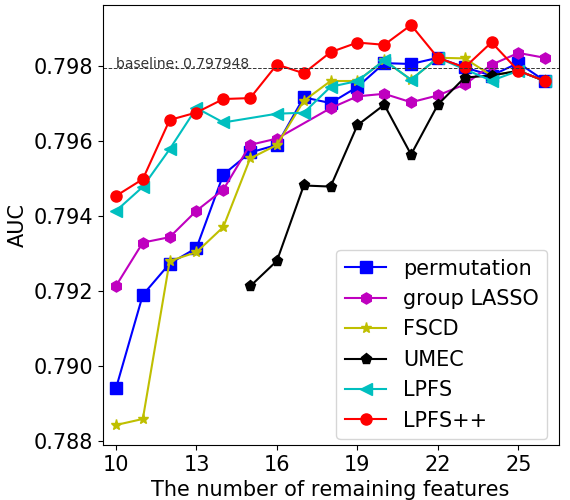}
\end{minipage}
\caption{Performance comparison of our method with other methods. Where the abscissa is large(i.e. only a few features are removed), there is little difference between all these methods. We find in experiments that the feature subsets selected by these methods are almost the same. However, where the abscissa is small (i.e. around half of the features are removed), our method has clear superiority over other methods. The baseline for 27 features is best AUC 0.797948, best ACC 0.811079, best Loss 0.423315.}
\label{fig: terabyte}
\end{figure}

The result is shown in Figure~\ref{fig: terabyte}. When only a small number of features are supposed to be removed, we find in experiments that all these methods select almost the same subsets, so there is little difference in AUC obtained by these methods (look at the upper right corner of Figure~\ref{fig: terabyte}). While when we want to remove about a half, where the number of combinations becomes much larger($C_{27}^{14}\approx 2\times10^7$), our method can cope with this challenge well, so as to select the most informative subset (look at the left half of Figure~\ref{fig: terabyte}). The feature subset our method selects is attached in the supplementary material for future reference. Although we compare AUC rather than ACC here for UMEC, the ACC for UMEC in our experiment is still higher than the result in the original paper under the same number of remaining features. In the following experiments for a large feature sets, we do not compare with UMEC, partly because of its poor performance here and partly because it is difficult to compress to a very small subset.

\subsection{Terabyte With Cross Features}
\label{Terabyte With Cross Features}
\begin{figure}
\centering
\begin{minipage}[t]{0.45\textwidth}
\centering
\includegraphics[width = 0.85\textwidth]{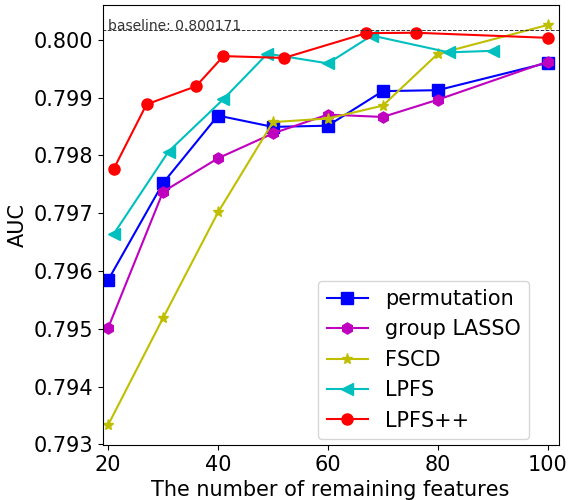}
\end{minipage}
\caption{Performance comparison for Terabyte with cross features. Every two different feature embeddings are crossed by element-wise multiplication. Then the original first-order features and the crossed second-order features are concatenated as the input of the \textsl{top}-MLP of DLRM. So, the total number of features can be viewed as $27+C_{27}^2=378$. We can see that our methods, LPFS and LPFS++, have many advantages over other methods, especially when the number of remaining features is small. It can be seen that our methods can also be used to mine higher-order features.}
\label{fig: terabyte_cross}
\end{figure}

In this experiment, we take the experiment a step further, partly because 27 features were so few, and partly because we want to show that our method can be easy to be applied to study feature interactions. In this experiment, every two embeddings corresponding to two different feature fields are crossed by element-wise multiplication, then the original features and the crossed features are concatenated as the input of the \textsl{top}-MLP of DLRM. So, the total number of features can be viewed as $27+C_{27}^2=378$, and we treat these 378 features equally to select from this much larger superset. So, if an original feature should be removed, the features that interact with it are not necessarily removed. Except for the cross-feature and the resulting larger number of input units for the first fully connected layer of \textsl{top}-MLP of DLRM, other experiment configurations are the same as the last subsection. In this way, we can both select first and second-order features. The result is shown in Figure~\ref{fig: terabyte_cross}. Compared with previous works~\cite{2019AutoCross,liu2020autofis}, our methods can also be used to select higher-order features effectively.

\subsection{Industrial Dataset}
\begin{figure}
\centering
\begin{minipage}[t]{0.45\textwidth}
\centering
\includegraphics[width = 0.85\textwidth]{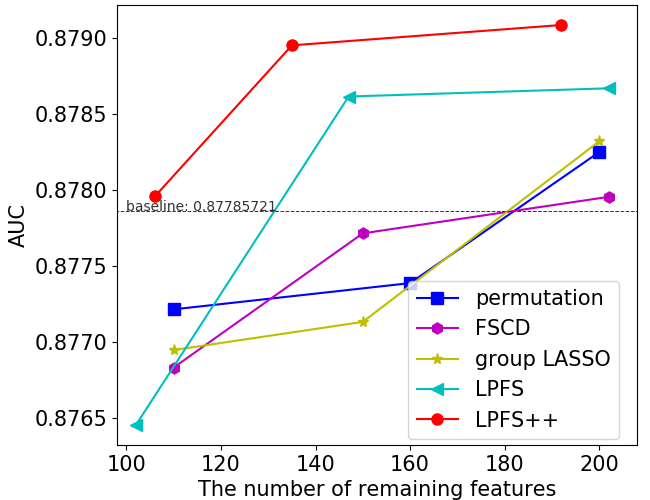}
\end{minipage}
\caption{Performance comparison of our methods with other methods for KuaiShou industrial dataset. The total number of feature field is 392 (250 user, 46 item, 96 combine), and we treat all these features equally. In the network, every user feature and every item feature is interacted by element-wise multiplication, and there are five auxiliary tasks to help improve the performance of the main task. Our LPFS++ has more obvious advantages in industrial large-scale datasets and challenging network structure.}
\label{fig: biaodan}
\end{figure}

In this experiment, we apply our methods to a large-scale industrial dataset. All the 392 categorical feature fields (250 users, 46 items, 96 combined) mentioned above are treated equally for the feature selection, although the network does not treat these features equally. Due to the existence of cross features between user and item, even if the same number of features is kept, the computational amount may not be the same. Still, we only care about the feature subsets, rather than the computation cost. Different from the experiment~\ref{Terabyte With Cross Features}, we insert gates immediately after the embeddings, rather than after the crossed features. So, when a feature is supposed to be removed, all cross-features that interact with it will also be removed. The result is shown in Figure~\ref{fig: biaodan}. As can be seen from the figure, our method has more obvious advantages in industrial large-scale datasets and challenging complex network structures.
\begin{figure}
\centering
\begin{minipage}[t]{0.45\textwidth}
\centering
\includegraphics[width = 0.85\textwidth]{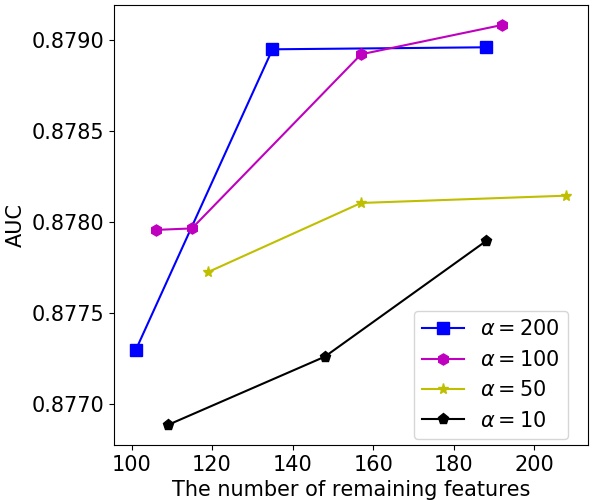}
\end{minipage}
\caption{Ablation study for different $\alpha$. We can see that in general, the larger the $\alpha$ value, the more informative the selected feature subset. }
\label{fig:ablation}
\end{figure}
\subsection{Ablation Studies and Experimental Analysis}
The influence of $\epsilon's$ decaying rate on performance has been studied by~\cite{guo2021gdp} for channel pruning, and similar to it, the decaying rate of $\epsilon$ in ~\eqref{smoothL0} and ~\eqref{eq:SL0pp} is not much important for offline feature selection, and our experiments found that the final value of $\epsilon$ between $1e-4$ and $1e-8$ has no significant effect on the performance of the model. The key component of the gate function~\eqref{eq:SL0pp} for LPFS++ is the second arctangent function part and its balancing factor $\alpha$. For simplicity, we will focus on $\alpha$.

By Function~\eqref{SL0ppgredient}, since the derivative at $x=0$ is $g'_{\epsilon++}(x=0)=\alpha {\epsilon}^{1/\tau}$, we should fix the schedule of $\epsilon$ first before studying $\alpha$. In the industrial experiments, we decay $\epsilon$ by $0.986$ every 500 steps, and the minimal value is $1.0e-4$. As shown in Figure~\ref{fig:ablation}, in general, the larger the $\alpha$ value, the more informative the selected feature subset. This is to be expected, a larger value of $\alpha$ results in a larger value of the derivative $g'_{\epsilon++}(x=0)=\alpha {\epsilon}^{1/\tau}$ and thus a greater fault tolerance of the model. We also find in experiments that a larger $\alpha$ will result in a smaller number of features \revise{to be removed.}
%in the selected feature subset, 
when other hyper-parameters kept unchanged. Therefore, in order to select roughly the same number of features, when increasing the value of $\alpha$, the value of $\lambda$ in Function~\eqref{eq:objective} must also be increased accordingly, as shown in Table~\ref{alphalambda}. This is also to be expected, because a larger value of $\alpha$ makes the model more resistant to compression. \\
\begin{table}
	\centering
% 	\begin{center}
	\setlength{\tabcolsep}{3mm}{
 	\begin{tabular}{ccccc}
		\toprule
		$\alpha$ & 10 & 50 & 100 & 200\\
		\midrule
        $\lambda$ & 0.004 & 0.0066 & 0.013 & 0.02\\
        \midrule
        \# feats & 109 & 119 & 106 & 101\\
		\bottomrule
	\end{tabular}
	}
% 	\end{center}	
	\caption{The table shows the relationship between $\alpha$ and $\lambda$ when about the same number of features are wanted to be kept. "\# feats" means the number of remaining features. We can see that when we increase $\alpha$, the $\lambda$ should also be increase if we want to remain about the same number of features.}
	\label{alphalambda}
\end{table}
\begin{figure}
\centering
\begin{minipage}[t]{0.45\textwidth}
\centering
\includegraphics[width = 0.85\textwidth]{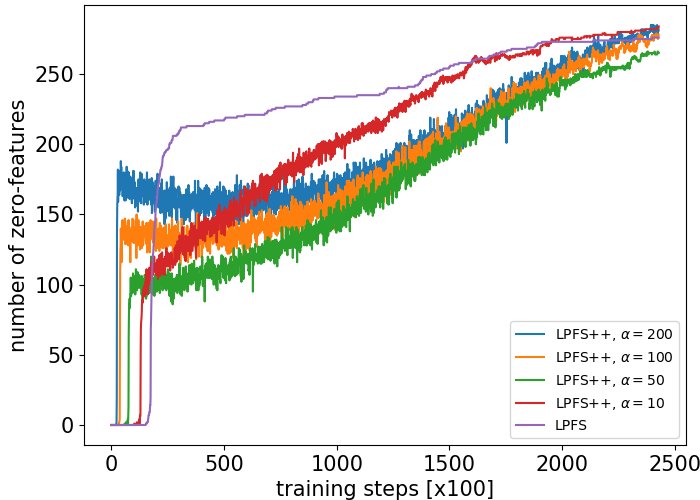}
\end{minipage}
\caption{The number of features with zero-gate changes over training step. The X-axis represents training steps (we record every 100 steps, so the total number of training steps is about 250000), while the Y-axis represents the number of features with zero-gate.}
\label{fig: biaodangandiao}
\end{figure}
Figure~\ref{fig: biaodangandiao} shows the number of features with zero-gate changes over the training step, and it contains a wealth of interesting information. First of all, let's look at the beginning of the training, where zero-gates first appear (i.e. the ordinate of the curve becomes non-zero for the first time). In our experiments, all the $x$ in Function~\eqref{smoothL0} and ~\eqref{eq:SL0pp} are initialized to $1.0$. So, at the beginning of training, the first case (i.e. $\tilde{x}_t \ge 2\lambda\eta$) in ~\eqref{eq: closed-form prox-l1} dominates, and $x$ will be subtracted by $2\lambda\eta$ at each step. Then as the training goes on, $x$ becomes zero gradually, and the speed at which it becomes zero is related to the value of $2\lambda\eta$. In Figure~\ref{fig: biaodangandiao}, we tune the $\alpha$ and $\lambda$ so that the final number of removed features is about the same. \revise{And it shows that zero gates occur earlier in the model with larger $\lambda$ and $\alpha$.}
%And it shows that the model with larger $\lambda$ and $\alpha$ will get zero gates earlier. \\
Second, in the middle of training, as the value of $\alpha$ increases, the oscillating degree of the curve also increases. This is also as expected, since increasing $\alpha$ increases the $g'_{\epsilon++}(x=0)=\alpha {\epsilon}^{1/\tau}$, giving more features a chance to revive.\\
Third, near the end of the training, all the curves gradually stop oscillating, because $g'_{\epsilon++}(x=0)=\alpha {\epsilon}^{1/\tau}$ and the learning rate is decaying.
\subsection{Online Experiments}
The proposed LPFS and LPFS++ have been successfully deployed on the offline Kuaishou distributed training platform and exploited by dozens of advertising scenarios in Kuaishou.
To show the effectiveness of our proposed methods, we choose our first launch as an example, in which nothing changed except for the input feature fields.
The dataset is collected from KuaiShou APP and the feature superset contains $332$ feature fields due to the extensive feature engineering, from which we select $177$ features for subsequent experiments. \revise{We found that the feature subsets selected across different runs are highly overlapping, which shows that our method is robust.}
We first conduct an offline experiment on the dataset which is sampled from $14$ days online logs.
Our model relatively outperforms permutation feature importance and the previous online feature set by $0.209\%$ and $0.304\%$ in offline AUC, respectively.
We conducted an online A/B test for two weeks in July 2021.
Compared with the previous state-of-the-art method used online, our method increased the cumulative revenue \cite{du2021exploration} by $1.058\%$ and the social welfare \cite{du2021exploration} by $1.046\%$, during the A/B test of 10\% traffic.
% In this subsection, we take an online experiment for the single-column item-impression task from KuaiShou as an example to show our method. The feature superset contains over 300 feature fields, then we select 177 from it. The dataset is collected from KuaiShou APP, and we pre-train the model by "May 10$\sim$ May 24", then search features by "May 21$\sim$ May 24". Before getting online, we evaluate the 177 features offline first: we train a model by cold start for 7 days then evaluate on 1 day, then get AUC 0.8841 (compared to AUC 0.8820 via feature shuffle, AUC 0.8811 of base model). After getting online, we used 10\% of our traffic to observe the performance from June 24 to June 30. The cost for the information flow market index increased by 1.744\%, and the customer expected expenditure increased by 0.79\%.

% \section{Appendix}

\subsection{Complexity Analysis}
As complexity analysis is important for industrial recommender systems, we do a simple complexity analysis in this section. Suppose there are total $N$ features, we need to initialize a $N$-dimensional learnable vector $\boldsymbol{x}$ in equation~\eqref{network}, which consumes $4N$ bytes storage. Then the intermediate variables include $g(\boldsymbol{x})$, as well as the gradient and the momentum of $\boldsymbol{x}$ during the optimization process, approximately occupying a total of $4\times4N$ bytes memory, which is negligible compared with the storage of the whole network and the embeddings. Moreover, we found in many experiments that the increase of the training time in each iteration is also negligible.

In our experiments, if the training of deep CTR model with the feature superset is cold-started, it needs to be pre-trained for a while until almost converged, using $T_1$ time, and we then load this pre-trained checkpoint to perform feature superset pruning to obtain the best feature subset, using $T_2$ time. Then we find that $T_2$ is usually about $\frac{1}{4}T_1$ in our all experiments. Of course, if we do feature selection based on an online model checkpoint, this pre-training step can be omitted for this case.

\subsection{Experiment details}
In order to facilitate readers to reproduce and use our method, we describe our experiment in detail.

In all the experiments, $x$ in Function~\eqref{eq:SL0pp} and ~\eqref{smoothL0} is initialized to $1.0$, while $\epsilon$ $0.1$. It can be calculated that the initial value of Function~\eqref{eq:SL0pp} and ~\eqref{smoothL0} is $0.909$ and $0.909+0.785\alpha\epsilon$. When we get a pre-trained model, in order to ensure that there is no sudden change in the performance of the model before and after inserting the gates, we divide all gate functions by their initial value, so that the initial value of the gate function is 1.0. Besides, if we only do feature selection for categorical features, rather than continuous features, the magnitude of categorical feature embeddings will change quickly compared with that of continuous representations, and this phenomenon is especially evident for LPFS++. So, we find that it would be better to divide the categorical feature embedding by an overall number, which can be the root mean square of gate values. This number does not participate in gradient backpropagation, and can be updated every 100 steps for example, then keep fixed after enough training steps. We found that this step resulted in a slight improvement in LPFS++ performance.

For Terabyte, we did not shuffle the dataset, and we train in chronological order to sense changes in user behavior to show the robustness of LPFS++. We use Adagrad optimizer for the model parameters (not including the gate parameters), and the batch size is 512, and the learning rate is 0.01. We decay $\epsilon$ by 0.9978 every 100 steps, and the minimum value is $1.e-5$. We use proximal-SGD with momentum to train $x$ in Function~\eqref{eq:SL0pp} and ~\eqref{smoothL0}, the initial learning rate is $0.01$ for LPFS++, $0.005$ for LPFS and decayed by 0.9991 every 100 steps, but not exceed $5e-4$. Different $\alpha$ and $\lambda$ values are combined to tune the parameters to obtain feature subsets of different sizes. 

For the industrial dataset, we also use Adagrad optimizer for the model parameters (not including the gate parameters), and the batch size is 1024, learning rate is initial as $0.01$ then exponentially rises to $0.1$ very slowly and stays constant. We decay $\epsilon$ by $0.986$ every $500$ steps, with minimal value $1e-4$.

\section{CONCLUSION, LIMITATION and FUTURE WORK}
In this paper, we adopted the idea of smoothed-$\ell^0$ to feature selection and proposed a new smoothed-$\ell^0$-liked function to select features more effectively and robustly. Both LPFS and LPFS++ show superiority over other methods, and LPFS++ achieves state-of-the-art performance. LPFS and LPFS++ have played an important role as efficient feature selection plugin tools for recommendation scenarios in Kuaishou Technology.

The main limitation for function~\eqref{eq:SL0pp}, from the perspective of dimensional analysis, is that it is not scale-invariant. The $x$ in the arctangent trigonometric function $\arctan$ means that $x$ must be dimensionless. For function~\eqref{smoothL0}, $\epsilon$ has the dimension of $x$ squared. If we want to construct a gate function that is just a dimensionless coefficient, then the derivative must have the dimension of the form $\sim \frac{1}{x}$ or $\sim \frac{x}{\epsilon}$. When $x=0$ and $\epsilon$ is small enough, the $\sim \frac{1}{x}$ tends to be infinity, $\sim \frac{x}{\epsilon}$ tends to be strictly zero, and it is impossible to have finite non-zero derivative values at $x=0$. Thus we have to assume that both $x$ and $\epsilon$ are dimensionless, and the consequence that function~\eqref{eq:SL0pp} is not scale-invariant is inevitable. Obviously, the solution is not unique to the problem, and there are many smoothed-$\ell^0$-liked functions that satisfy the derivative being non-zero and decaying as $\epsilon$ decaying. Other than function~\eqref{eq:SL0pp}, we didn't try any other similar property functions yet.

In fact, LPFS++ can be easily extended to online feature selection. What we want is that some features can be active or inactive dynamically as user behavior changes, and then predict whether a user will click on an item using active features in real-time. Our LPFS++ can meet well with this requirement. Specially, we can maintain a superset of features to train and select features simultaneously, so that all the model parameters and gate parameters are trained together online. Then, the features in the active status and their corresponding sub-networks can become effective on a specialized inference platform in real-time. Online feature selection enables the model to capture the temporal changes in user behaviors and select the optimal feature subset dynamically and adaptively in real-time. This involves both algorithmic, training, and inference engineering improvements, which are left as future work.
%%
%% The acknowledgments section is defined using the "acks" environment
%% (and NOT an unnumbered section). This ensures the proper
%% identification of the section in the article metadata, and the
%% consistent spelling of the heading.
% \begin{acks}
% To Robert, for the bagels and for explaining CMYK and color spaces.
% \end{acks}
%\clearpage
%%
%% The next two lines define the bibliography style to be used, and
%% the bibliography file.

%%
%% If your work has an appendix, this is the place to put it.
%\clearpage
%\newpage

\appendix
% \section{Supplementary materials}
\section{Appendix}
\subsection{Feature mask for Terabyte dataset}
We list the feature subsets that we selected by LPFS++ for reference. In the following table~\ref{feature mask}, the "\#" column is the number of remaining features, the "mask" column indicates the feature subsets selection. In the mask, "0" means the corresponding feature should be removed, while "1" means kept. The length of every mask is $27$, where the first value represents the $13$ continuous features, and the following $26$ values represent the $26$ categorical feature fields. The order of the feature fields is the same as the official code of DLRM~\cite{naumov2019deep}.
\begin{table}[h]
\small
	\centering
	\setlength{\tabcolsep}{0.7mm}{
 	\begin{tabular}{cc|cc}
		\toprule
		\# & mask & \# & mask \\
		\hline
		26 & 111110111111111111111111111 & 17 & 111100010111110101010101011\\
		\hline
        25 & 111100111111111111111111111& 16 & 111100010111110001010101011\\
        \hline
        24 & 111100111111111111011111111& 15 & 111100010011110101000101011\\
                \hline
        23 & 111100011111111111011111111& 14 & 111100010011110001000101011\\
                \hline
        22 & 111100011111111101011111111& 13 & 111100010011110001000001011\\
                \hline
        21 & 111100110111110101011111111& 12 & 111100010011110000000001011\\
                \hline
        20 & 111100010111110101011111111& 11 & 111100010101110000000001001\\
                \hline
        19 & 111100010111110101011101111& 10 & 101100010100110000010101000\\
                \hline
        18 & 111100010111110101011101011& & \\
		\bottomrule
	\end{tabular}
	}
	\caption{Feature mask for Terabyte dataset.}
	\label{feature mask}
\end{table}

\subsection{Core code}
\begin{figure}[h]
\centering
\includegraphics[width = 0.49\textwidth]{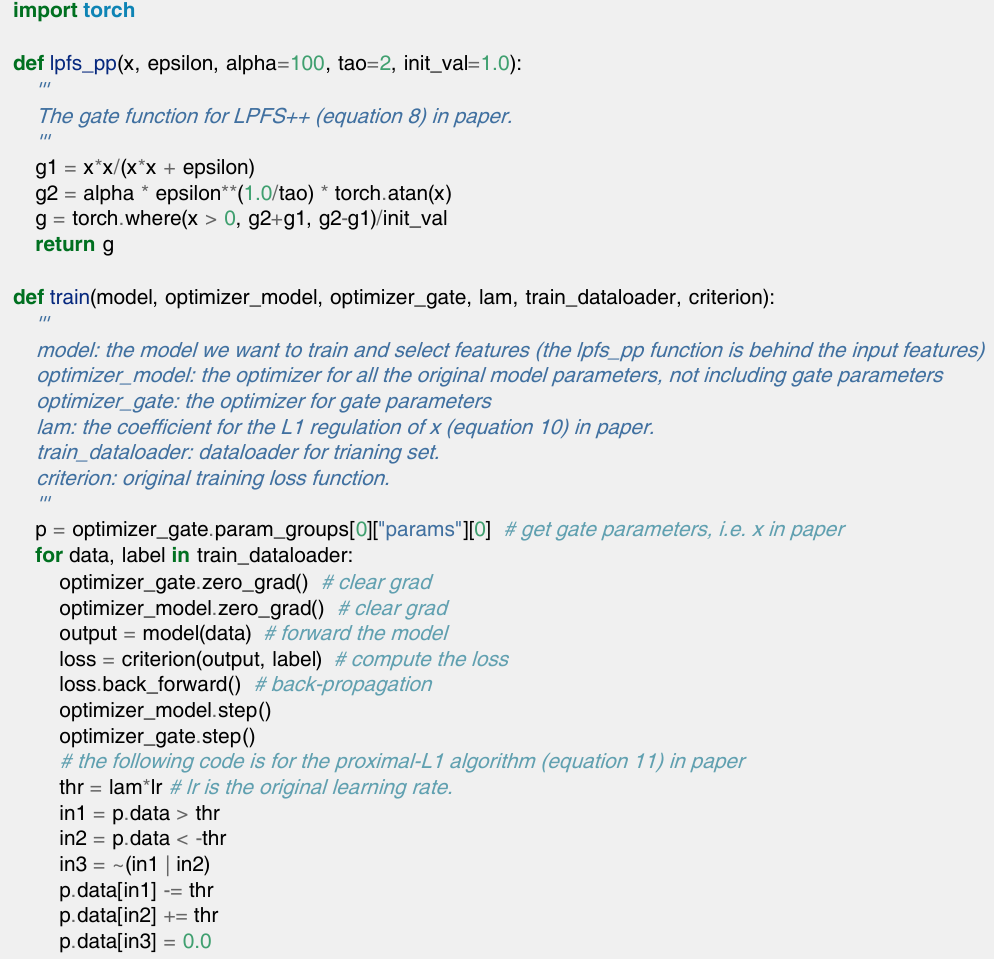}
\caption{Our LPFS++ can be implemented within several lines of code using PyTorch. To demonstrate the reproducibility, we list the core code of LPFS++ and the proximal-SGD updating here.}
\label{fig:code}
\end{figure}

\clearpage
\bibliographystyle{ACM-Reference-Format}
\bibliography{sample-base}

\end{document}